# Experiments and Modeling of the Autoignition of Methylcyclohexane at High Pressure

Bryan W. Weber*[a], William J. Pitz[b], Marco Mehl[b], Emma Silke[b], Alexander C. Davis[c], Chih-Jen Sung[a]

[a]Department of Mechanical Engineering, University of Connecticut, Storrs, CT, USA

[b]Lawrence Livermore National Laboratories, Livermore, CA, USA

[c]Clean Combustion Research Centre, King Abdullah University of Science and Technology, Thuwal, Saudi Arabia

*Corresponding Author:
Email: bryan.weber@uconn.edu
Address: 191 Auditorium Road U3139, Storrs, CT, 06269
Phone: 860-486-2492

**Experiments and Modeling of the Autoignition of Methylcyclohexane at High Pressure**

Bryan W. Weber[1], William J. Pitz[2], Marco Mehl[2], Emma Silke[2], Alexander C. Davis[3], Chih-Jen Sung[1]

[1]Department of Mechanical Engineering, University of Connecticut, Storrs, CT, USA

[2]Lawrence Livermore National Laboratories, Livermore, CA, USA

[3]Clean Combustion Research Centre, King Abdullah University of Science and Technology, Thuwal, Saudi Arabia

**Abstract**

New experimental data are collected for methyl-cyclohexane (MCH) autoignition in a heated rapid compression machine (RCM). Three mixtures of MCH/$O_2$/$N_2$/Ar at equivalence ratios of $\phi$=0.5, 1.0, and 1.5 are studied and the ignition delays are measured at compressed pressure of 50 bar and for compressed temperatures in the range of 690–900 K. By keeping the fuel mole fraction in the mixture constant, the order of reactivity, in terms of inverse ignition delay, is measured to be $\phi$=0.5 > $\phi$=1.0 > $\phi$=1.5, demonstrating the dependence of the ignition delay on oxygen concentration. In addition, an existing model for the combustion of MCH is updated with new reaction rates and pathways, including substantial updates to the low-temperature chemistry. The new model shows good agreement with the overall ignition delays measured in this study, as well as the ignition delays measured previously in the literature using RCMs and shock tubes. This model therefore represents a strong improvement compared to the previous version, which uniformly over-predicted the ignition delays. Chemical kinetic analyses of the updated mechanism are also conducted to help understand the fuel decomposition pathways and the reactions controlling the ignition. Combined, these results and analyses suggest that further investigation of several of the low-temperature fuel decomposition pathways is required.

# 1. Introduction

Cycloalkanes and alkyl-cycloalkanes (collectively known as naphthenes) are well known major components in several transportation fuels, including gasoline, diesel, and jet fuels [1–4]. Since these transportation fuels have hundreds or thousands of individual chemical components, incorporation of all these components in a kinetic model would make the model very difficult to build and computationally expensive to use. To facilitate modeling such real fuels, it is necessary to formulate a surrogate mixture by selectively choosing a much smaller set of neat components that will reproduce the physical and chemical behavior of the target fuel. Methyl-cyclohexane (MCH) is frequently suggested as a candidate component in these formulations to represent the naphthene content of real fuels [5,6]. Furthermore, an understanding of MCH kinetics can provide the base from which to build models of the combustion of other naphthenes.

Many studies have been conducted to examine the combustion behavior of naphthenes, and such an extensive literature review is beyond the scope of this work; we therefore limit our focus in the following to MCH and the conditions of particular interest to this study. The interested reader is directed to the review of Pitz and Mueller [7] for a comprehensive discussion of experimental and modeling work relevant to naphthenes.

If chemical kinetic models are to be used in engine design, it is critical that they are able to reproduce the combustion behavior of fuels under the thermodynamic conditions prevalent in engines. Furthermore, they must be of an appropriate size to enable calculations on a reasonable time scale. So-called "reduced mechanisms" suitable for design calculations are typically validated by comparison to the detailed mechanisms from which they are derived. Thus, it is critical that the detailed mechanism first be able to reproduce the combustion properties of a given fuel under the desired conditions.

New engines, using advanced concepts such as homogeneous charge compression ignition (HCCI) and reactivity controlled compression ignition (RCCI), incorporate Low Temperature Combustion (LTC) to help achieve the goals of improved fuel efficiency and lower emissions. However, a detailed understanding of LTC reaction pathways is often required to properly predict the combustion phasing, heat release rates, and engine-out emissions of such engine concepts. In addition, since HCCI and RCCI engines operate at high pressures and their combustion performance is sensitive to fuel chemistry, the chemical kinetic models used in engine simulations need to be validated at these conditions. Therefore, experimental data acquired at engine-relevant conditions are of critical importance for validating chemical kinetic model performance.

One of the most common global validation parameters for kinetic models is the ignition delay. The ignition delay is especially relevant to reciprocating engines, as it helps determine the combustion phasing and heat release rate of a fuel. Ignition delays of MCH have been measured in shock tubes [8–13] and rapid compression machines (RCMs) [14–16] by a number of researchers. These studies collectively cover the temperature-pressure space in the range of 700–2100 K and 1–70 atm. To complement this experimental work, a number of kinetic models for MCH combustion have been constructed, notably by Orme et al. [12] and Pitz et al. [15].

However, the existing models are not able to predict ignition delays at conditions for which they were not validated (i.e. the models are not truly predictive). For instance, previous work conducted in an RCM by Mittal and Sung [16] measured the ignition delays of MCH/$O_2$/$N_2$/Ar mixtures at pressures of 15.1 and 25.5 bar, for three equivalence ratios of $\phi$=0.5, 1.0, and 1.5, and over the temperature range of 680–840 K. They compared their measured ignition delays to simulated ignition delays computed using the mechanism of Pitz et al. [15] and found that the model over-predicted both the first stage and overall ignition delay substantially [16].

In view of the above, the objectives of this work are twofold. The first objective is to collect new autoignition data in an RCM at a higher pressure of 50 bar and over a temperature range that includes the LTC range. The datasets include the pressure history that relates to the heat release rates in the RCM, the first stage ignition delay time that is characteristic of the low temperature combustion, and the total ignition delay time that corresponds to hot ignition in an engine. The second objective of this paper is to update the reaction pathways and rate constants of important reactions in the kinetic model of Pitz et al. [15] and use the previously and newly obtained RCM data to validate the updated model. By comparing the experimental and simulated ignition delay results and conducting chemical kinetic analyses of the updated mechanism, the fuel decomposition pathways and the controlling reactions for autoignition are identified and discussed.

## 2. Methods

### 2.1. Quantum Chemical Calculations

Since $RO_2$ isomerization rate constants were available for cyclohexane but not for methylcyclohexane, ab initio calculations were performed at LLNL by the authors for a particularly important $RO_2$ isomerization that involves an abstraction from the methyl group. Specifically, the isomerization is for the 2-methylcyclohexyl-1-peroxy radical, in which the –OO group is attached at the beta position relative to the methyl group. This reaction was shown to potentially have a large rate constant because of its low activation energy, as calculated by Yang et al. [17]. Because we needed the pre-exponential factor that was not reported in the work by Yang et al. [17], we calculated the high-pressure rate constant using the CBS-QB3 composite method [18]. The reactant, product, and transition state geometries and frequencies were calculated using the Gaussian09 suite of programs [19]. The lowest energy conformations were obtained from relaxed scans around each rotor in 60° increments using B3LYP/6-31+G(d,p). Reaction rates and Eckart tunneling factors were determined using ChemRate [20]. The transition state was confirmed by the presence of a single imaginary frequency that corresponds to the reaction pathway and IRC calculations using B3LYP/6-31+G(d,p). The reaction rate was determined using a rigid rotor harmonic oscillator approximation with corrections for hindered rotors. Hindered rotors were accounted for using the Pitzer and Gwinn 1D hindered rotor method within ChemRate, with the rotational barriers being determined via relaxed scans in 10° increments using B3LYP/6-31+G(d,p) [21]. The high-pressure rate-limited reaction rate was fit to a three-parameter Arrhenius rate expression over the temperature range of 300–2500 K.

## 2.2. Experimental Methods

The experimental facility consists of a rapid compression machine, a mixture preparation system, and diagnostics. For mixture preparation, the fuel and oxidizer pre-mixtures are prepared in a stainless steel mixing tank. The volume of the tank is approximately 17 L so that many experiments can be run from a single batch. The liquid fuel (methyl-cyclohexane, 99.0% purity) is massed to a precision of 0.01 g in a syringe before being injected into the mixing tank through a septum. The proportions of oxygen (99.9999% purity), nitrogen (99.9995% purity), and argon (99.9999% purity) are determined by specifying the oxidizer composition, the equivalence ratio, and the total mass of fuel. The gases are added to the mixing tank manometrically at room temperature.

The mixing tank, reaction chamber, and all lines connecting them are equipped with heaters to prevent condensation of the fuel. The partial pressure of the fuel is kept below the saturation vapor pressure for all preheat temperatures and equivalence ratios studied. After filling the tank, the heaters are turned on and the system is allowed approximately 1.5 hours to equilibrate. The mixture is stirred continuously by a magnetic vane to ensure mixture homogeneity. This procedure has been validated in several previous studies [22–24]. In these studies, the concentrations of n-butanol, n-decane, and water were verified by GCMS, GC-FID, and GC-TCD, respectively, to be within 5% of the expected value.

Three different mixtures of MCH/$O_2$/$N_2$/Ar are prepared in this study, as outlined in Table 1. These mixtures (denoted as Mix #1–3) match the mixtures prepared in our previous work with MCH in the RCM [16]. The equivalence ratios corresponding to Mix #1–3 are $\phi$=1.0, 0.5, and 1.5, respectively. As in the previous RCM experiments, the mole fraction of MCH is held constant and the mole fraction of $O_2$ is varied to adjust the equivalence ratio. This experimental design allows these data to be used to validate chemical kinetic models for changes in $O_2$ concentration, which is an important variable in internal combustion engines where exhaust gas recirculation is used to reduce the oxygen concentrations to avoid NOx formation. Few validation data for ignition are available for changing oxygen concentrations. In addition, the relative proportions of $O_2$, $N_2$, and Ar are adjusted so that the same specific heat ratio is maintained in the three mixtures. The utility of this experimental design will be discussed in due course.

Table 1: Molar Proportions of Reactants

| Mix # | $\phi$ | MCH | O$_2$ | N$_2$ | Ar |
|---|---|---|---|---|---|
| 1 | 1.0 | 1 | 10.5 | 12.25 | 71.75 |
| 2 | 0.5 | 1 | 21.0 | 0.00 | 73.50 |
| 3 | 1.5 | 1 | 7.0 | 16.35 | 71.15 |

The RCM used for these experiments is a pneumatically-driven/hydraulically-stopped single-piston arrangement and has been described in detail previously [25]. At the start of an experimental run, the piston rod is held in the retracted position by hydraulic pressure while the reaction chamber is vacuumed to less than one torr. Then, the reaction chamber is filled with the required initial pressure of test gas mixture from the mixing tank. The compression is triggered by releasing the hydraulic pressure. The piston assembly is driven forward to compress the test mixture by high-pressure nitrogen gas. The gases in the test section are brought to the compressed pressure ($P_C$) and compressed temperature ($T_C$) conditions in approximately 30 milliseconds. The piston in the reaction chamber is machined with a specifically designed crevice to ensure that the roll-up vortex effect is suppressed and homogeneous conditions in the reaction chamber are promoted.

The homogenous conditions in the reaction chamber allow the assumption that the reactants are compressed nearly adiabatically [25]. Therefore, $P_C$ and $T_C$ are only a function of the temperature-dependent specific heat ratio of the reactants, the compression ratio, and the initial conditions. In the present study design, the specific heat ratio of the reactants is held constant across the three equivalence ratios, as mentioned previously. Thus, for given $P_C$, compression ratio, and initial conditions, the $T_C$ will be the same for all the equivalence ratios.

In the present operation procedure, $P_C$ and $T_C$ can be varied independently by adjusting the volumetric compression ratio and the initial temperature ($T_0$) and initial pressure ($P_0$) of the test charge. The pressure in the reaction chamber is monitored during and after compression by a rapid-response, thermal-shock resistant piezoelectric dynamic pressure transducer (Kistler 6125B). During the filling of the mixing tank and reaction chamber prior to compression, the pressure is monitored by a static pressure transducer.

Figure 1 shows a representative pressure trace from these experiments at $P_C$=50 bar, $T_C$=761 K, and $\phi$=1.5 (Mix #3). The definitions of the end of compression (EOC) and the ignition delays are indicated on the figure. The end of compression time is defined as the time when the pressure reaches its maximum before first stage ignition occurs, or for cases where there is no first stage ignition, the maximum pressure before the overall ignition occurs. The first stage ignition delay is the time from the end of compression until the first peak in the time derivative of the pressure. The overall ignition delay is the time from the end of compression until the largest peak in the time derivative of the pressure.

Figure 1 also shows a non-reactive pressure trace. Due to heat loss from the test mixture to the cold reactor walls, the pressure and temperature of the gas in the reaction chamber will decrease after the end of compression. A non-reactive pressure trace is measured that corresponds to each unique $P_C$ and $T_C$ condition studied to quantify the effect of the heat loss on the ignition process and to verify that no heat release has occurred during the compression stroke. The non-reactive pressure trace is acquired by replacing the oxygen in the oxidizer with nitrogen, so that the specific heat ratio of the initial mixture is maintained, but the heat release due to exothermic oxidation reactions is eliminated. Maintaining a similar specific heat ratio ensures that the non-reactive experiment faithfully reproduces the conditions of the reactive experiment. A representative non-reactive pressure trace is shown in Fig. 1 corresponding to the experimental conditions in the figure.

Each unique $P_C$ and $T_C$ condition is repeated at least 6 times to ensure repeatability of the experiments. The experiment closest to the mean of the runs at a particular condition is chosen for analysis and presentation. The standard deviation of all of the runs at a condition is less than 10% of the mean in all cases. Furthermore, to ensure reproducibility, each new mixture preparation is checked against a previously run condition before new data is collected using that mixture.

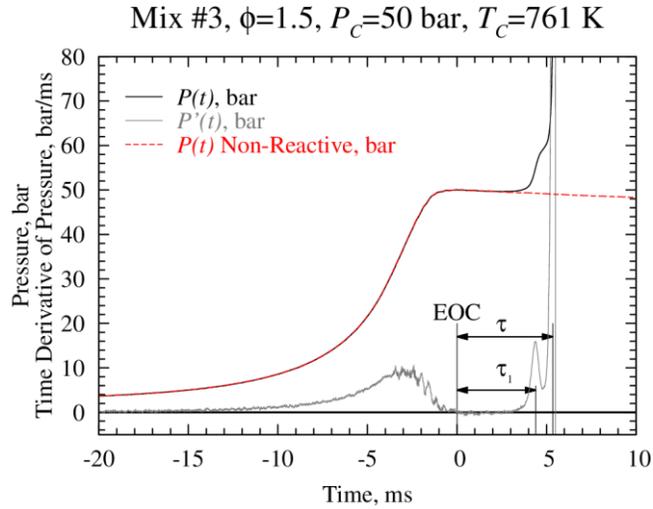

Figure 1: Representative pressure trace indicating the definition of the first stage ($\tau_1$) and overall ($\tau$) ignition delays and the corresponding non-reactive pressure trace. EOC stands for End of Compression.

## 2.3. Numerical Methods

In simulations, a volume history is used to account for the effects of the compression stroke and heat loss in the experiments. These volume histories are derived from the reactive and non-reactive pressure traces collected for each unique $P_C$ and $T_C$. This procedure uses the temperature-dependent specific heat ratio and assumes an isentropic process; the compression stroke is modeled as an isentropic compression and the heat loss after EOC is modeled as an isentropic volume expansion. Using the isentropic process relations, the volume as a function of time is calculated during and after the compression stroke. Two types of simulations are then conducted using the Closed Homogeneous Batch Reactor in CHEMKIN-Pro 15131 [26]. The first type accounts for the effect of the compression stroke and post-compression heat loss by tabulating the reactor volume function of time using the volume trace computed by the isentropic relations discussed previously. The tabulated volume is referred to as the volume profile and is specified in the CHEMKIN-Pro input file with the VPRO keyword. Therefore, this type of simulation is referred to as a VPRO simulation hereafter. The second type is a constant volume, adiabatic simulation and is referred to as CONV hereafter, again named after the CHEMKIN-Pro input keyword used to specify this problem type. The CONV type simulations do not capture the effect of the compression stroke and post-compression heat loss; therefore, they allow direct analysis of the kinetic model with no influence of potentially confounding experimental effects.

VPRO simulations are used to calculate the temperature at the end of compression, $T_C$. This temperature is used as the reference temperature for reporting the ignition delay. This approach requires the assumption of a homogeneous, adiabatic core of gases in the reaction chamber, which is facilitated on the present RCM by the creviced piston described previously. Simulations to determine $T_C$ are conducted with and without detailed reaction steps to determine if there is significant reactivity in the compression stroke. If there is no significant reactivity (and hence heat release), the pressure and temperature at EOC are the same whether or not reactions are included in the simulation.

## 3. Model Development

The LLNL MCH chemical kinetic model has been updated to reflect new chemical kinetic information that has become available since the publication of our 2007 mechanism [15]. The following sub-models in the mechanism have been replaced: the $C_1$-$C_4$ base chemistry with the AramcoMech version 1.3 [27]; the aromatics base chemistry with the latest LLNL-NUIG model [28]; and the cyclohexane sub-model with a more recent version from Silke et al. [29]. In addition, there are several specific updates of the MCH mechanism. The abstraction reactions from MCH are replaced using recent experimentally measured values [30] and standardized using the latest LLNL reaction rate rules [31]. The previous 2007 MCH model [15] lumped many of unsaturated ring products of MCH, including methylcyclohexenes and methylcylcohexadienes. These species have now been expanded to include all the relevant isomers with their associated reaction paths and rate constants. The model now tracks the intermediate unsaturated methyl-cyclohexane species with much more fidelity and predicts their experimentally measured concentrations with much more accuracy [32].

Regarding the low temperature chemistry portion of the chemical kinetic mechanism, there have been many updates. For the R+$O_2$ reactions involving the cyclohexane ring in MCH, the ab initio rate constants computed by Fernandes et al. [33] were used. They computed the rate constants from 1 to 50 bar over a temperature range of 500–900 K. Since $RO_2$ isomerization rate constants were available for cyclohexane but not for the case of a methyl substitution on the ring, ab initio calculations were performed at LLNL by the authors for the case of a six membered ring where the –OO group is attached at the beta position relative to the methyl group. These calculations are discussed in separate sections of this paper. Rate constants for $R+O_2$ involving a tertiary site were not available from Fernandes et al. [33], so these rate constants were estimated based on the rate constants of their secondary site counterparts. The A-factors were adjusted for degeneracy since there is only one tertiary C-H and the activation energy was reduced by 2 kcal for a tertiary abstraction rather than a secondary abstraction. Since pressure dependent rate constants were not available for all the low temperature reactions for MCH (e.g. $R+O_2$, $RO_2$ isomerization), we used the high-pressure rate constants for all of them for consistency. For the rate constants obtained from Fernandes et al. [33] where the high-pressure rate constant was not reported, we used the 50 bar rate constant. Villano and Dean [34] found that the use of the high-pressure limit for $RO_2$ isomerization rates worked well for n-butane when the pressure was 10 atm and above. Since the pressures of interest in this study are 15 bar and above, we expect this to be a good assumption here as well. We also consider this a good assumption for pressures encountered in internal combustion engines.

Cyclic ether formation in cycloalkanes requires special considerations compare to acyclic alkanes. 4-membered ring cyclic ethers are observed in the low temperature oxidation of acyclic alkanes and are formed from $\gamma$-QOOH (the radical site is $\gamma$ from the OOH group) though a 4-membered ring transition state. However, Gulati and Walker [35] theorized that the chair structure of cyclohexane is too rigid to allow the formation of the four-membered ring cyclic ether and the formation of hexenal is instead more energetically favored. This is supported by the experimentally observed absence of 4-membered ring cyclic ethers and the detection of alkenones in the oxidation of cycloalkanes. Specifically, Gulati and Walker [35] observed hexenal, but not 4-membered ring cyclic ether in the oxidation of cyclohexane. Husson et al. [36] similarly observed octenone in the oxidation of ethylcyclohexane in a jet stirred reactor and but not the corresponding 4-membered ring cyclic ether. In the present mechanism, the $\gamma$-QOOH forms heptenone using the rate constant computed by Fernandes et al. [33] at 50 bar for the

cyclohexyl+$O_2$ system. Because of the lack of information on their oxidation, different isomers of heptenone have been lumped into one species in the mechanism.

One of the changes to the mechanism was to increase the activation energy of the ketohydroperoxide decomposition from 39 kcal/mol (163.2 kJ/mol) (Reaction Class 28 in [31]) to 41.6 kcal/mol (174.1 kJ/mol). 41.6 kcal/mol is the activation energy used for $C_4$ ketohydroperoxide decomposition in the updated $C_1$-$C_4$ base chemistry from Metcalfe et al. [27]. This increase in activation energy was made to improve the agreement between the measured and simulated ignition delay times (as shown later), to be consistent with the base chemistry, and to move toward an activation energy that is closer to the O-O bond strength of the ketohydroperoxide (44.5 kcal based on group additivity estimates). The simulated ignition delay time is very sensitive to this decomposition rate constant, which is largely controlled by the bond strength. Fundamental experiments and calculations are needed to obtain a more accurate estimate of this rate constant.

For the thermodynamic parameters for species, we have adopted those in AramcoMech [27] for $C_1$-$C_4$ species. For other new species in the mechanism such as the unsaturated cyclic species derived from MCH, THERM [37] was used to calculate the thermodynamic parameters.

The chemical kinetic mechanism and the thermodynamic database are available as supplemental data in CHEMKIN format. The reaction mechanism is fully annotated with details on the pedigree of rate constants not covered here. The current model contains 6498 reactions among 1540 species.

## 4. Results

### 4.1. Quantum Chemical Calculations

In their work on the low temperature oxidation of cycloalkanes, Yang et al. [17] reported the effect of methyl substitution on the viability of 1,4 and 1,5 H-migration reactions. Of importance to the current study was their evaluation of the barrier heights of the 1,4 and 1,5 H-migration of the cis- and trans- forms of 2-methylcyclohexyl-1-peroxy radicals (c2McHP and t2McHP, respectively). Due to the preference of the equatorial arrangement of methylcyclohexane compared to the axial one, they only reported the barrier heights involving the former. Yang et al. [17] reported barrier heights of 123.7, 120.5, and 92.5 kJ/mol for the 1,5 H-migrations in c2McHP for reactions involving the hydrogen on $C_5$ and $C_3$ on the ring and the methyl group, respectively, while for t2McHP only the methyl site is available for a 1,5 H-migration and has a barrier height of 96.2 kJ/mol. Unfortunately, for our purposes, Yang et al. [17] did not report full rate expressions. As a result, we repeated their calculations for their lowest energy pathway, the c2McHP 1,5 H-migration involving the migration of the methyl group hydrogen. While performing a conformational analysis, a lower energy structure was obtained for c2McHP (Fig. 2). By reducing the energy of the starting point, the barrier height increases by approximately 2.5 kJ/mol, since both pathways go through the same transition state. Taking into account the three-fold degeneracy of the hydrogens at the methyl site, a rate of $k(T) = 1.92 \times 10^6 T^{1.81} e^{-82.5/RT}$ s$^{-1}$, where the activation energy is in kJ/mol and $R$ is the universal gas constant, is obtained for the 1,5 H-migration, which is similar to, but slightly faster than, similar reactions in linear alkylperoxy radicals [38].

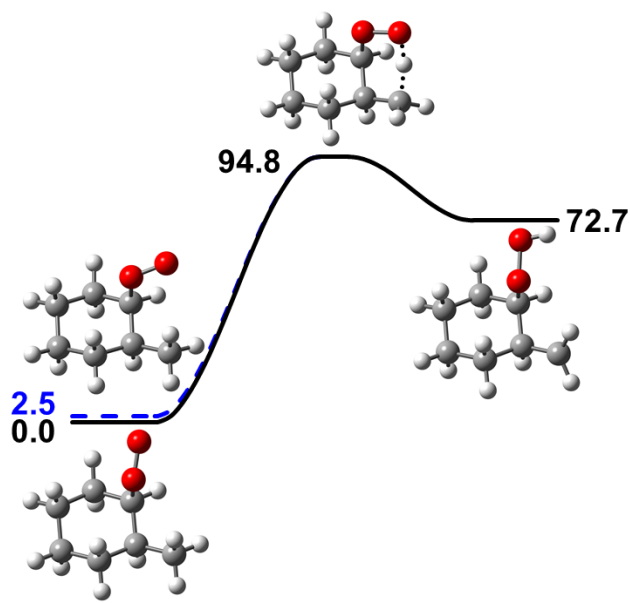

**Figure 2: Reaction path diagram for the 1,5 H-migration reaction in 2-methylcyclohexyl-1-peroxy (c2McHp) radical at 298 K. Energies are in kJ/mol. Blue dashed line represents pathway described in Yang et al. [17].**

## 4.2. Experimental Results

The experimental ignition delays measured at the three equivalence ratios and compressed pressure of 50 bar are shown in Fig. 3. The open symbols are the overall ignition delays and the filled symbols are the first stage ignition delays. The vertical error bars on the experimental data represent twice the standard deviation of all of the experiments at that condition. Detailed uncertainty analysis of the deduced compressed temperature was conducted previously by Weber et al. [22], who performed a detailed error propagation analysis considering the uncertainty in the initial pressure, initial temperature, initial mixture composition, and compressed pressure measurements. They found that the maximum uncertainty of the compressed temperature was approximately $\pm 1\%$ for their RCM experiments; since the same RCM is used in this study, we adopt the same temperature uncertainty.

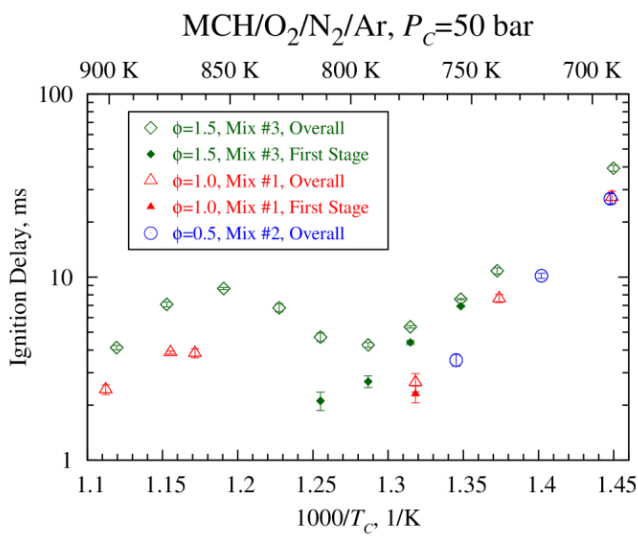

Figure 3: Experimentally measured ignition delays at $P_C$=50 bar for the mixture conditions in Table 1.

The negative temperature coefficient (NTC) region is an important feature of low temperature ignition where the ignition delay time increases with increasing temperature. The NTC region of the overall ignition delay is evident in Fig. 3 for the $\phi$=1.5 case (Mix #3) and approximately includes the temperature range of $T_C$=775–840 K. For $\phi$=1.5, first stage ignition is evident for conditions in the range of $T_C$=740–800 K.

For $\phi$=1.0 (Mix #1), the NTC region of the overall ignition delay could not be completely resolved. Only three conditions in the low temperature region and three conditions in the high temperature region are shown in Fig. 3. The experimental pressure traces during the compression stroke for intermediate temperature conditions were seen to deviate from their non-reactive counterparts, demonstrating appreciable reactivity therein. Hence, those data are not included in Fig. 3.

For the experiments at $\phi$=0.5 (Mix #2), only three data points in the low temperature region are reported and none of them exhibits two-stage ignition response. As the temperature is increased further, noticeable reactivity during the compression stroke is evident.

As stated earlier, the mole fraction of MCH is held constant in this study, while the mole fraction of the oxidizer is changed to modify the equivalence ratio. Figure 3 demonstrates that the $\phi$=0.5 case is the most reactive (as judged by the inverse of the ignition delay) and the $\phi$=1.5 case is the least reactive. As has been shown for other fuels, including *n*-butanol [22] and Jet-A [39], decreasing the equivalence ratio by increasing the oxygen mole fraction but holding the fuel mole fraction constant increases the reactivity.

### 4.3. Comparison to Model

A comparison of the experimentally measured first stage ignition delays (open symbols) and the first stage ignition delays computed using the updated model (lines) is shown in Figs. 4(a), 5(a), and 6(a) for Mix #1, #2, and #3, respectively. In addition, a comparison of the experimentally measured overall ignition delays (open symbols) and the overall ignition delay computed by the updated model (lines) is shown in Figs. 4(b), 5(b), and 6(b). The experiments include the new work being presented here at $P_C$=50 bar in addition to the previous RCM experiments at $P_C$=15.1 and 25.5 bar [16]. The simulations are the VPRO type of simulations. For some computational cases, substantial heat release during the compression stroke caused the computed pressure to depart from the non-reactive profile prior to EOC. Therefore, these cases are not shown in Figs. 4–6. For these conditions, the experimental pressure trace did not exhibit significant heat release during the compression stroke and the experimental pressure at EOC for the reactive case matched that of the non-reactive counterpart. The volume traces provided to the VPRO simulations are available from the authors' website at http://combdiaglab.engr.uconn.edu/database/rcm-database. The ignition delays measured experimentally in this study are available in the Supplementary Material.

At 15.1 and 25.5 bar for Mix #1 and #2, the overall ignition delay is very well predicted for temperatures above approximately 715 K. For lower temperatures at these two equivalence ratios, the experimental ignition delays are under-predicted by the model, but the predictions are nevertheless within a factor of two of the data. For the rich case (Mix #3), the simulations under-predict the ignition delay over a wider temperature range but the results improve as temperature increases. Again, the experimental ignition delays are predicted to within approximately a factor of two. At 50 bar, the ignition delays are under-predicted for all of the equivalence ratios studied here, but the agreement is within a factor of two.

The first stage ignition delays for all of the pressure and equivalence ratios are under-predicted, but are within a factor of three of the experimental values. Furthermore, for all of the equivalence ratios tested at $P_C$=50 bar, it is of interest to note that there are several cases where simulated ignition delays show two-stage response where the experiment shows only a single stage ignition. Nevertheless, the present mechanism is a marked improvement from the comparison performed by Mittal and Sung [16] who found that the ignition delays were strongly and uniformly over-predicted by the previous LLNL mechanism by Pitz et al. [15].

Figures 7(a), 7(b), and 7(c) show a comparison of selected simulated and experimentally measured pressure traces for Mix #1, #2, and #3, respectively, at $P_C$=50 bar. Also shown in Fig. 7 is the simulated non-reactive pressure trace corresponding to each experimental condition. Small differences in the heat loss profile for different temperatures are apparent in the non-reactive pressure traces. These differences arise from the changing surface area to volume ratio of the reaction chamber at the end of compression as the compression ratio is changed to vary the compressed temperature. This highlights the importance of using VPRO simulations to compare predictions of ignition delay with the experimental data.

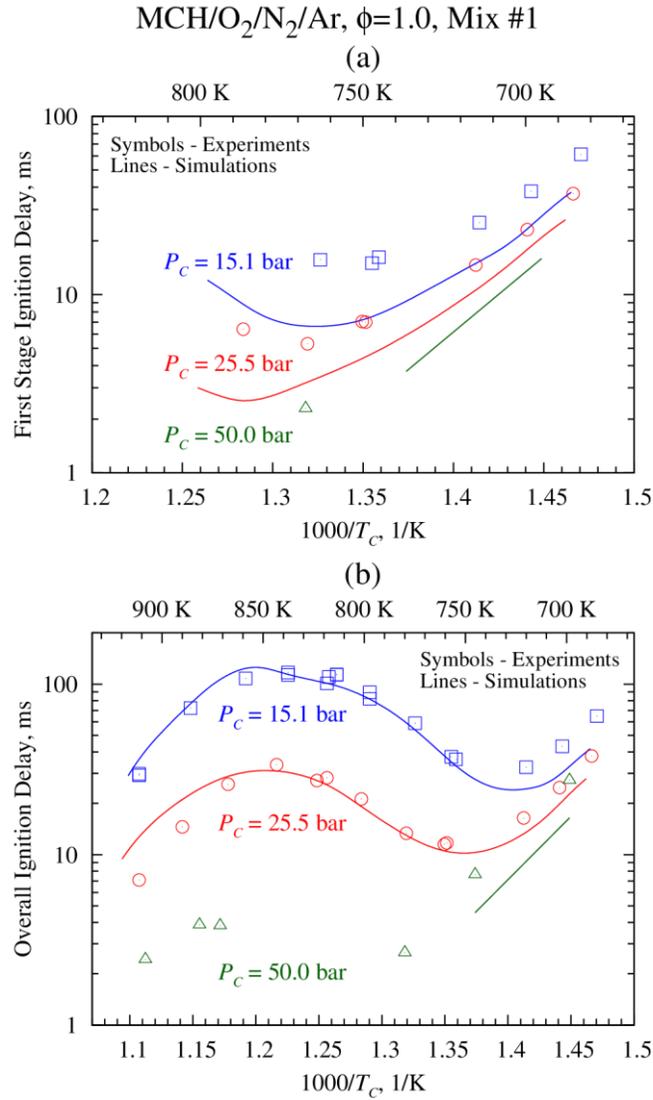

**Figure 4: Comparison of experimental and simulated ignition delays for three pressures for Mix #1. The data at 15.1 and 25.5 bar are from the study of Mittal and Sung [16]. (a) First stage ignition delays (b) Overall ignition delays.**

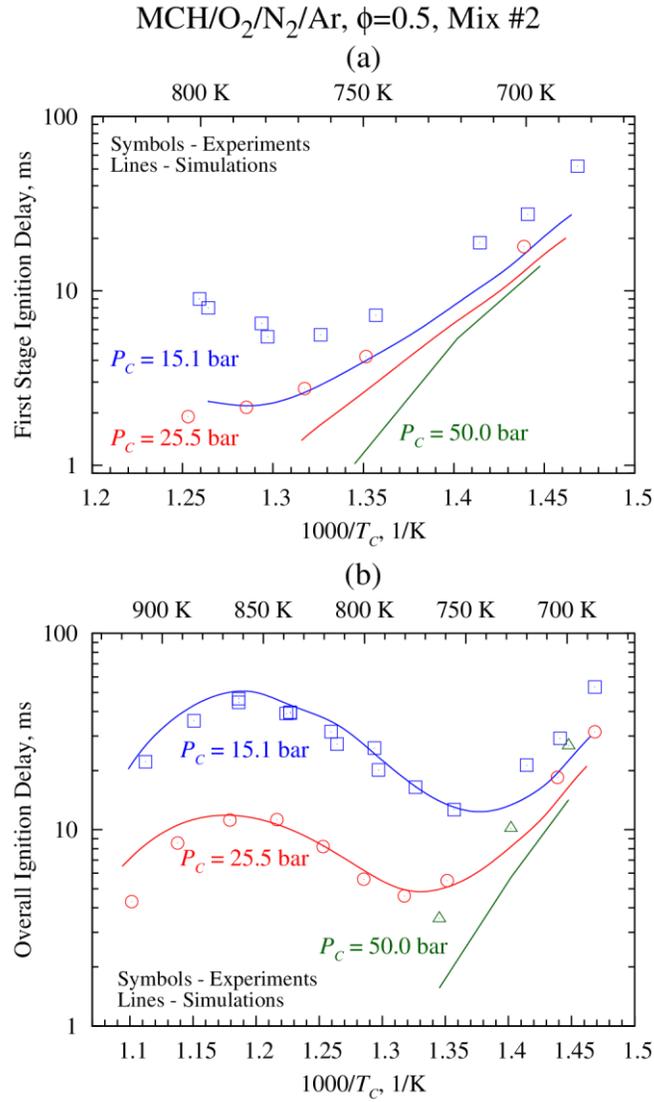

**Figure 5:** Comparison of experimental and simulated ignition delays for three pressures for Mix #2. The data at 15.1 and 25.5 bar are from the study of Mittal and Sung [16]. (a) First stage ignition delays (b) Overall ignition delays.

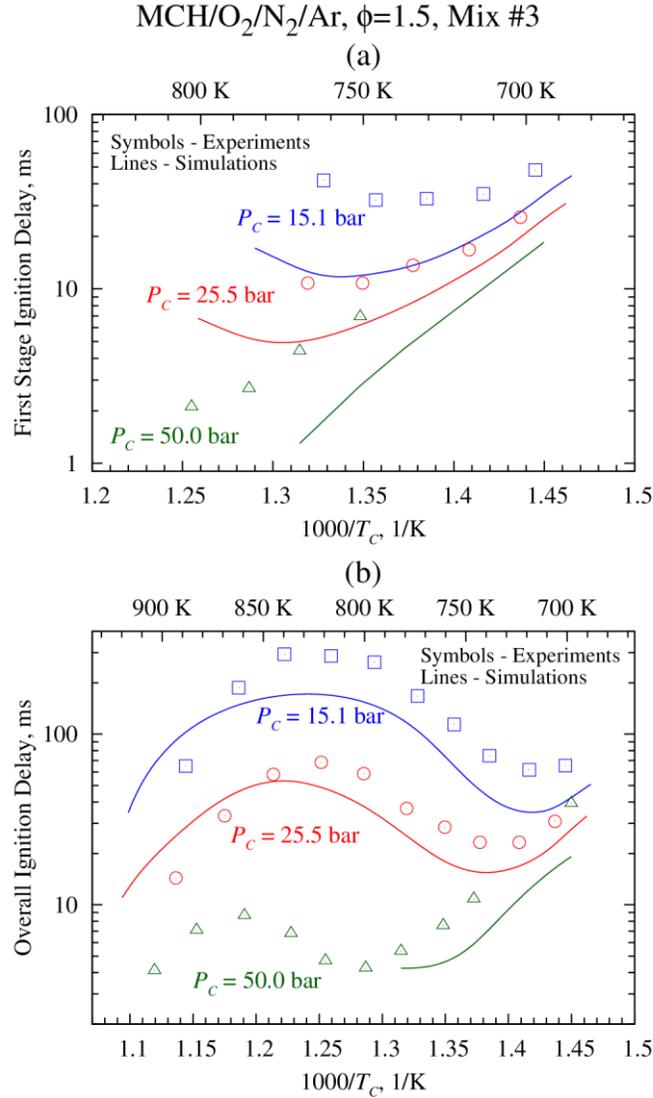

**Figure 6: Comparison of experimental and simulated ignition delays for three pressures for Mix #3. The data at 15.1 and 25.5 bar are from the study of Mittal and Sung [16]. (a) First stage ignition delays (b) Overall ignition delays.**

For Mix #1, it is clear that the simulated reactive pressure trace in Fig. 7(a) at $T_C$=866 K (red dashed line) deviates from the non-reactive pressure trace (red dot-dot-dashed line) prior to the end of compression. The same is also true of the 797 K case shown for Mix #3 in Fig. 7(c). Remarkably, the simulated case for Mix #1 at $T_C$=866 K predicts the overall ignition delay quite well. However, due to the heat release prior to EOC, this simulated result is not plotted in Fig. 4. The simulated case for Mix #3 at $T_C$=797 K is also not plotted on Fig. 6 due to the heat release prior to EOC; interestingly, this case under-predicts the first stage ignition delay but over-predicts the overall ignition delay. For the other simulated cases (black lines), the reactive pressure traces closely follow their non-reactive counterparts until the ignition event begins. The experimental ignition delays of these cases are under-predicted by the model. It is also seen in Fig. 7(c) for $T_C$=729 K that the model predicts two-stage ignition, although two-stage ignition is not observed experimentally.

The current mechanism is also compared to shock tube ignition delays from the studies of Vasu et al. [9] and Vanderover and Oehlschlaeger [10]. Those studies considered the autoignition of stoichiometric mixtures of MCH with $O_2/N_2$ air. The comparison is shown in Fig. 8 for the near 50 atm data from those studies. Note that the experimental data shown are the raw data and are not scaled to a constant pressure, whereas the simulated ignition delays are at a constant initial pressure of 50 atm. It can be seen that the ignition delays are over-predicted over the nearly entire temperature range of 795–1160 K studied. Nevertheless, the predicted ignition delays are within approximately a factor of 1.5 of the experiments, indicating good agreement overall and a substantial improvement from the previous version of the model. Furthermore, the simulations shown here are of the CONV type and do not account for any facility dependent effects present in the experiments. Although the experimentalists noted in [9,10] that the effect of such considerations is minimal in their studies, including facility dependent effects will tend to make the simulations ignite sooner and improve the agreement, especially for cases with ignition delays longer than approximately 1000 µs.

As discussed in Section 3, one of the updates to the model was to increase the activation energy of Reaction Class 28, ketohydroperoxide decomposition, from $E_a$=39 kcal/mol (163.2 kJ/mol) to 41.6 kcal/mol (174.1 kJ/mol). This update substantially improved the prediction of the low-temperature ignition delays, including the first stage and overall ignition delays. As mentioned by Curran et al. [40], "the high activation energy [of Reaction Class 28] ensures an induction period during which the ketohydroperoxide concentration builds up." Furthermore, updating this activation energy does not affect the high-temperature ignition delays. A comparison of calculated ignition delays demonstrating the effect of this update is shown in Fig. 9.

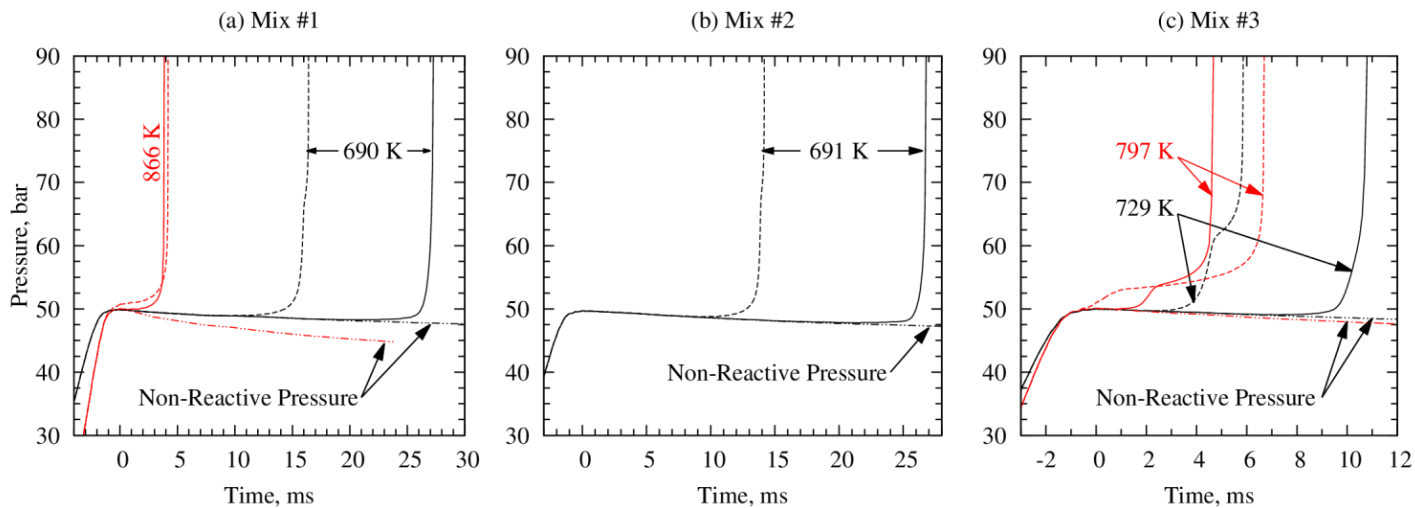

**Figure 7: Comparison of selected simulated and experimental pressure traces at $P_C$=50 bar for (a) Mix #1 (b) Mix #2 (c) Mix #3. Red lines indicate that the pressure profile of the reactive simulation deviates from the non-reactive case prior to EOC. Solid lines: experiment; dashed lines: reactive simulation; dot-dot-dashed lines: non-reactive simulation.**

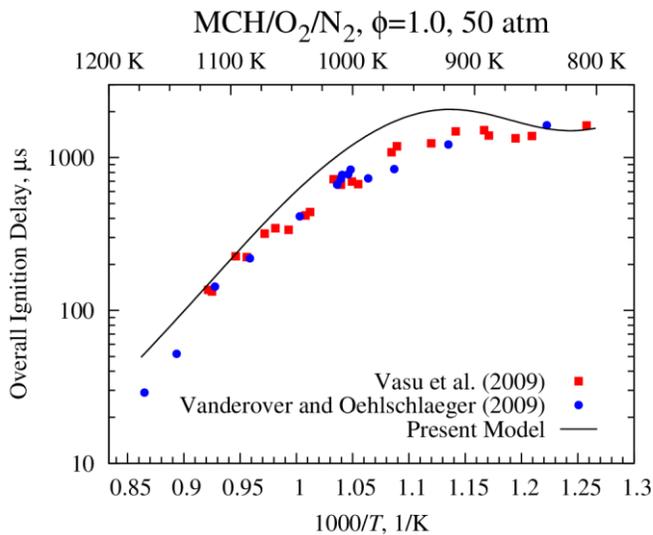

**Figure 8: Comparison of the present model with the experiments from Vasu et al. [9] and Vanderover and Oehlschlaeger [10] near 50 atm and for stoichiometric mixtures in $O_2/N_2$ air.**

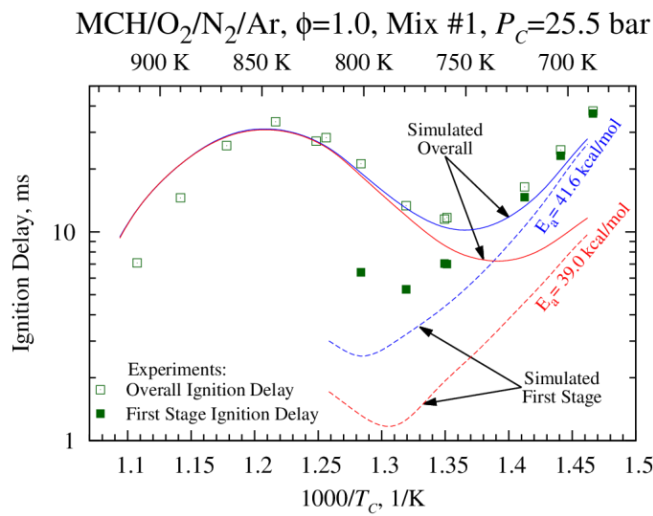

**Figure 9:** Comparison of mechanism performance with the activation energy of Reaction Class 28, ketohydroperoxide decomposition, set at 41.6 kcal/mol (blue) and 39.0 kcal/mol (red). Experimental ignition delays are shown in green symbols.

## 5. Discussion

### 5.1. Path Analysis

The relatively good agreement of the updated model with the experimental data suggests that a more detailed analysis of the mechanism is a worthwhile exercise and such analysis may point the way to further improvements to the mechanism. We begin with a reaction path analysis. The present reaction path analysis is conducted using a CONV (adiabatic, constant-volume) type simulation for three initial temperatures (700 K, 800 K, and 900 K), at 25.5 bar and for Mix #1 (the stoichiometric case). For the other mixture conditions and pressures considered in this work, the absolute percentages for each channel change slightly. However, the analysis of the reaction pathways is the same for all of the equivalence ratios and pressures considered in the experiments presented previously. The three temperatures considered in this analysis correspond to the low-temperature, peak of the NTC, and high-temperature portions of the ignition delay curve illustrated in Fig. 9; their results are shown in Fig. 10 with plain text, bold text, and italic text, respectively. The path analysis presented in Fig. 10 is an integrated analysis where the rate of production (ROP) of each species by each reaction has been integrated with respect to time up to 20% fuel consumption. The integrated ROPs from each reaction are normalized by the total production or destruction of that species up to 20% fuel decomposition, such that reactions that produce a species are normalized by the total production of the species and reactions that consume a species are normalized by the total consumption of that species. The percentages in Fig. 10 therefore represent the percent of the given reactant that is consumed to form the given product by all reactions that can form a particular product. Species such as hydroperoxyalkyl radicals (QOOH), alkyl hydroperoxides (ROOH), and methylcyclohexenes (MCH-ene) are shown as lumped on the path diagram; however, these species are un-lumped in the mechanism and presented as a lumped sum for simplicity in this diagram. Note that not all of the pathways present in the mechanism for each species are presented in Fig. 10, again for simplicity; the pathways that are shown in Fig. 10 typically account for more than 95% of the consumption of each species.

The first step of fuel breakdown occurs by H-atom abstraction at these pressure and temperature conditions. None of the fuel is directly decomposed by unimolecular reactions. Each of the seven possible radicals are formed in comparable quantities; however, due to the symmetry of MCH, sites 2 and 3 are equivalent to sites 6 and 5, respectively, so mchr2 and mchr3 have close to double the production rate compared to the other radicals. It is interesting to note that the production of mchr2, mchr3, and mchr4 increase as the initial temperature increases and the production of mchr1 and cychexch2 decrease to compensate. However, the change is small, no more than 2 percentage points for each radical.

The most important second step is oxygen addition (i.e. formation of ROO) at all of the initial temperatures in this analysis. The importance of this reaction diminishes for each radical as the initial temperature increases due to the increasing importance of β-scission reactions. At 700 K, less than 0.05% of each of the fuel radicals is consumed via β-scission. Between 800 and 900 K, the percentages of mchr1, mchr2, mchr3, and mchr4 that are decomposed via β-scission increase by several thousand percent each; nevertheless, the absolute change is small and the consumption of these radicals still occurs mostly by oxygen addition. The mchr1, mchr3, and mchr4 radicals undergo scission of the cyclohexyl ring, whereas mchr2 primarily scissions at the methyl-cyclohexyl bond. This beta scission of mchr2 competes significantly with its consumption by $O_2$ at 900 K. Furthermore, the increasing importance of the ring opening reactions from 800 K to 900 K means that chain propagation pathways (instead of effective chain termination pathways forming methylcyclohexene and hydroperoxyl) are available, increasing the reactivity. Finally, even at the elevated initial temperature of 900 K, cychexch2 does not undergo significant ring opening. Instead, it will scission an H atom from site 1 or steal an oxygen atom from hydroperoxyl to form an alkoxy radical (RO) when it does not undergo oxygen addition (these pathways each only consume about 0.3% of cychexch2 and hence are not shown in Fig. 10).

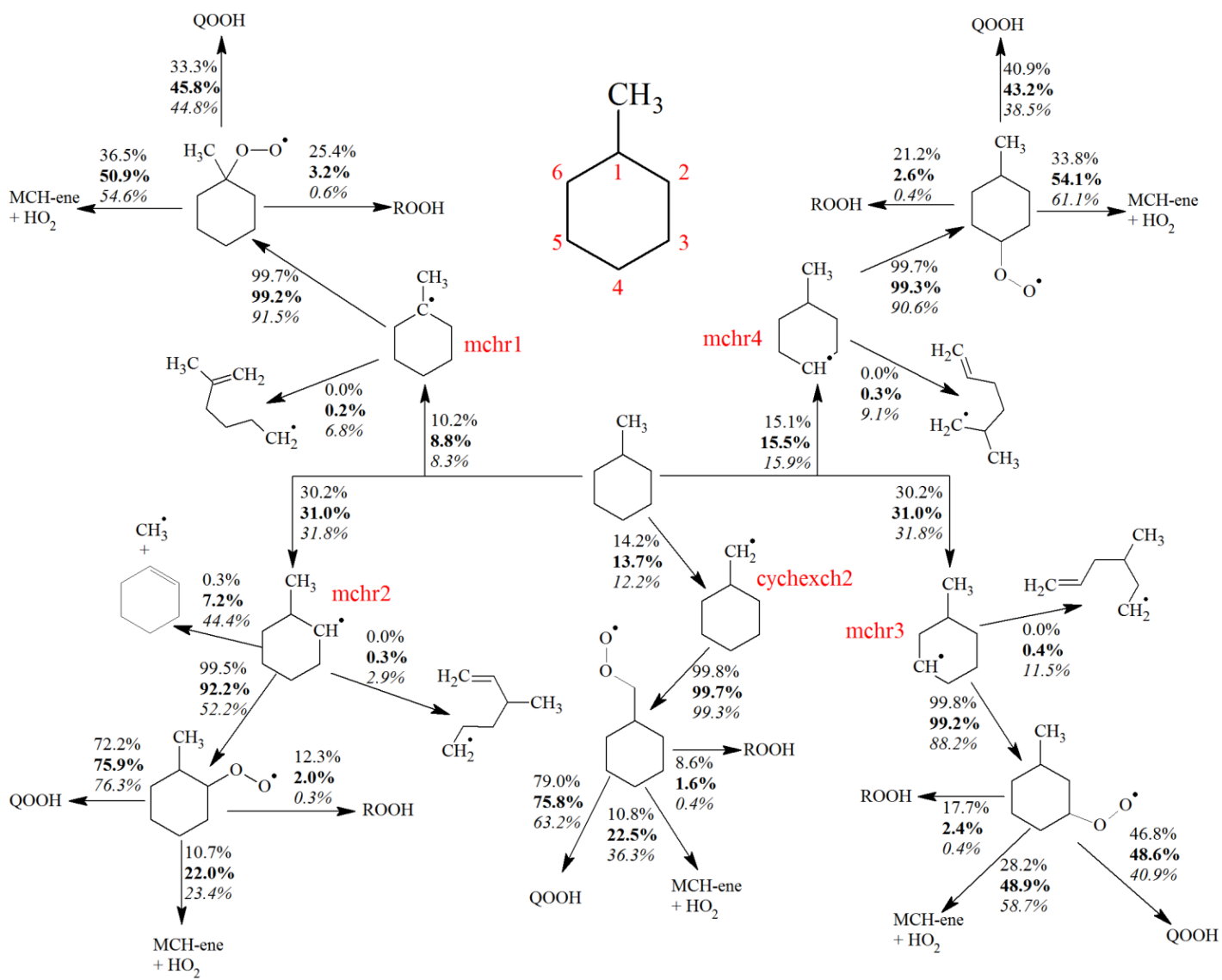

**Figure 10:** Path analysis of MCH combustion. Initial conditions are 25.5 bar and Mix #1 (ϕ=1.0) and 700 K (plain text), 800 K (bold text), 900 K (italic text). Note that not all possible reaction pathways are shown for each species.

Returning to the low temperature pathways, there are four important classes of reactions that consume the ROO radicals in the current mechanism. These classes are: C1) internal H-atom transfer (isomerization) to form QOOH; C2) direct elimination of hydroperoxyl and methylcyclohexene; C3) H-abstraction by ROO from either the fuel or hydroperoxyl to form ROOH; and, C4) reactions among the ROO radicals. Class C4 consumes less than ~5% of each of the ROO radicals at 700 K and less than ~0.1% for the other temperatures and this class is therefore not shown on the path diagram in Fig. 10. Of the other three classes, C1 (formation of QOOH) is the predominant pathway in the low temperature ignition process. Nevertheless, the direct elimination of methylcyclohexene and hydroperoxyl and the formation of ROOH are important at low temperatures as well.

For all of the temperatures considered here, a majority of the ROOH is formed by reactions of ROO with hydroperoxyl to give ROOH and an oxygen molecule. At the initial temperature of 700 K, approximately 15% of the fuel reacts to form ROOH, indicating its importance in low-temperature MCH combustion. The primary route of ROOH formation in this mechanism (H-abstraction from hydroperoxyl by ROO) has not been well studied at combustion relevant temperatures [41] and is therefore a good candidate for further investigation given its importance in the model for MCH combustion.

As the temperature increases, the formation of ROOH becomes substantially less important while the direct $HO_2$ elimination reaction becomes more important. The increase in production of methylcyclohexene and hydroperoxyl plays a role in the NTC region of ignition delay because this is effectively a chain terminating channel until the temperature increases enough that the sequence $MCH+HO_2=R+H_2O_2$; $H_2O_2(+M)=2OH(+M)$ becomes important and drives the overall ignition.

Interestingly, for most of the ROO radicals, the change in the fraction of ROO consumed to form QOOH is non-monotonic as temperature increases. That is, for mch1oo, mch3oo, and mch4oo the production of QOOH increases in going from 700 K to 800 K, then decreases going from 800 K to 900 K due to the increasing importance of the $HO_2$ elimination channel (due to nuances in the various reaction paths, mch2oo and chxch2oo do not follow this trend). Furthermore, the branching ratios in the decomposition of the QOOH species change as the temperature is increased (not shown in Fig. 10). At the lowest temperature (700 K), the formation of hydroperoxyalkylperoxy radicals (OOQOOH) is favored, leading to low-temperature chain branching and the two-stage ignition phenomenon. However, at 800 K and 900 K, the QOOH tends to decompose into a heptenone and a hydroxyl radical, or one of three epoxide species. In the current model, both heptenone and the three epoxide species are represented by lumped heptenone and epoxide species; however, due to the apparent importance of these species in the intermediate temperature decomposition of MCH, further investigation of their pathways is warranted.

### 5.2. Sensitivity Analysis

Our second type of analysis is a brute force, one-at-a-time sensitivity analysis. In this work, the sensitivity of the ignition delay to the reaction rates is considered. Due to the size of the mechanism, only the reactions of the fuel and the fuel radicals up to the OOQOOH species are considered. This approach is justified because many of the reactions of the $C_0$-$C_4$ base mechanism are known to be important to the ignition process (e.g., $H_2O_2(+M)=2OH(+M)$), but we are more interested in the effect of updates to the fuel specific sub-mechanism. The sensitivity index is defined in Eq. 1,

$$S_i = \frac{\ln(\tau_{i,2}/\tau_{i,1})}{\ln(k_{i,2}/k_{i,1})} \qquad \text{Equation 1}$$

where $\tau$ is the ignition delay time, either first stage or overall, $k$ is the reaction rate, and subscript $i$ indicates the reaction number. The numbered subscripts in Eq. 1 indicate the type of modification that has been made to the rate of reaction $i$ when computing the ignition delay, as discussed in the following.

The reaction rates are modified by multiplying and dividing the pre-exponential constant by a factor $f$. Thus, the forward and reverse rates are simultaneously modified. Special care is taken to properly modify reaction rates with pressure dependence and explicit reverse parameters. Each rate is modified sequentially and the ignition delay is computed; the pre-exponential constant is reset to its nominal value before modifying the next reaction. Finally, the nominal ignition delay with no rate modification is computed. Thus, each set of reactor input conditions requires $2N + 1$ model evaluations, where $N$ is the number of reactions considered in the sensitivity analysis and $N$ may be less than or equal to the total number of reactions.

The $2N + 1$ model evaluations result in $4N + 2$ ignition delays if two-stage ignition is present and $2N + 1$ ignition delays otherwise. These ignition delays are used to compute the sensitivity indices according to Eq. 1. In the case of bidirectional sensitivity indices, the subscript 2 in Eq. 1 is associated with multiplication by $f$ and the subscript 1 is associated with division by $f$, resulting in $2N$ sensitivity indices if two-stage ignition is present and $N$ indices otherwise. In the case of unidirectional sensitivity indices, the subscript 2 is associated with either multiplication or division by $f$ and the subscript 1 is associated with the nominal ignition delay, $\tau_{i,1} = \tau_1$. For unidirectional sensitivity indices, $4N$ indices are obtained if two-stage ignition is present and $2N$ are obtained otherwise.

In this work, the bidirectional sensitivity is used with $f=10$. For all of the reactions considered here, multiplying and dividing a given rate had opposite effects on the ignition delay. Thus, if the ignition delay increased (relative to the nominal case) when the rate of a certain reaction was multiplied, the ignition delay decreased (relative to the nominal case) when the rate of the same reaction was divided and vice versa. Since $k_{i,2}$ is greater than $k_{i,1}$ by definition, the sensitivity index $S_i$ will be positive if $\tau_{i,2} > \tau_{i,1}$ (i.e. increasing the rate increases the ignition delay) and negative if $\tau_{i,2} < \tau_{i,1}$ (i.e. increasing the rate decreases the ignition delay). The sensitivity analysis is run at the same conditions of the path analysis: CONV simulation, initial temperatures of 700 K, 800 K, and 900 K, initial pressure of 25.5 bar, and Mix #1. As with the path analysis, similar results are obtained for other pressures and mixtures.

Figure 11 shows the sensitivity indices for the five reactions (among all the reactions considered in the present sensitivity analysis) to which the overall ignition delay is most sensitive for each temperature studied (700 K, 800 K, and 900 K). For the results at 700 K and 800 K, the bidirectional sensitivity of the first stage ignition delay to the same reactions is also shown, except for two reactions at 800 K for which the unidirectional sensitivity is plotted. The reasons for this will be discussed in due course. It should be noted that the sensitivity indices of the first stage ignition delay have a slightly different ranking than the indices of the overall ignition delay. Therefore, the rank of the first stage sensitivity index of the reactions shown is given in parentheses next to the bar. At 700 K, the sensitivity of the overall ignition delay is in red and the sensitivity of the first stage ignition delay is in blue; at 800 K, the sensitivity of the overall ignition delay is in grey and the sensitivity of the first stage ignition delay is in green. The most sensitive reaction affecting the first stage ignition delay at 800 K is found to be MCH+OH=mchr3+$H_2$O, although it is not listed in Fig. 11. At 900 K, there is no first stage ignition, and thus no sensitivity of the first stage ignition delay.

Under the pressure/stoichiometry conditions of the present simulations, 800 K is approximately the highest initial temperature at which distinct two-stage ignition (i.e. two inflection points in the temperature or pressure trace) is found for MCH with the current mechanism. As such, several reactions affect the ignition strongly enough to eliminate the first inflection point. These reactions are given in Table 2 for either multiplication or division of the rate by the factor $f$=10. The naming convention of the species listed in Table 2 can be found in Fig. 10, Fig. 12, and the Supplementary Material. Two reactions shown in Table 2 also appear in Fig. 11, namely ($R1$) mch2oo=mch2ene+ho2 and ($R2$) mch2qx+o2=mch2qxqj. For these reactions at 800 K, the unidirectional sensitivity index is shown in Fig. 11, where $\tau_{i,2}$ in Eq. 1 is found by division of the rate for $i = R1$ and by multiplication of the rate for $i = R2$.

The role of the ROO=methylcyclohexene+$HO_2$ reactions in the left column of Table 2 in eliminating the first stage of ignition is clear – this set of reactions diverts ROO radicals from entering the low-temperature chain branching pathway via QOOH that leads to the two-stage ignition. Similarly, in the right column, decreasing the rate of the reaction of oxygen with QOOH to form OOQOOH reduces the rate of chain branching that leads to two-stage ignition. Concerning the reactions of the fuel with OH in the left column of Table 2, increasing these rates increases the formation of fuel radicals that are less reactive at low temperature than the cychexch2 and mchr2 radical. For example, the mchr2 radical adds to $O_2$ and forms a peroxy radical (mch2oo) that has a fast $RO_2$ isomerization path to QOOH involving the abstraction of an H atom from the methyl group. This $RO_2$ isomerization is the path calculated and discussed in Section 4.1. QOOH subsequently adds to O$_2$ and leads to chain branching. The high reactivity of cychexch2 and mchr2 at low temperature is reflected by the high percentages at 800 K (>70%) leading to QOOH from cychexch2oo and mch2oo in Fig. 10.

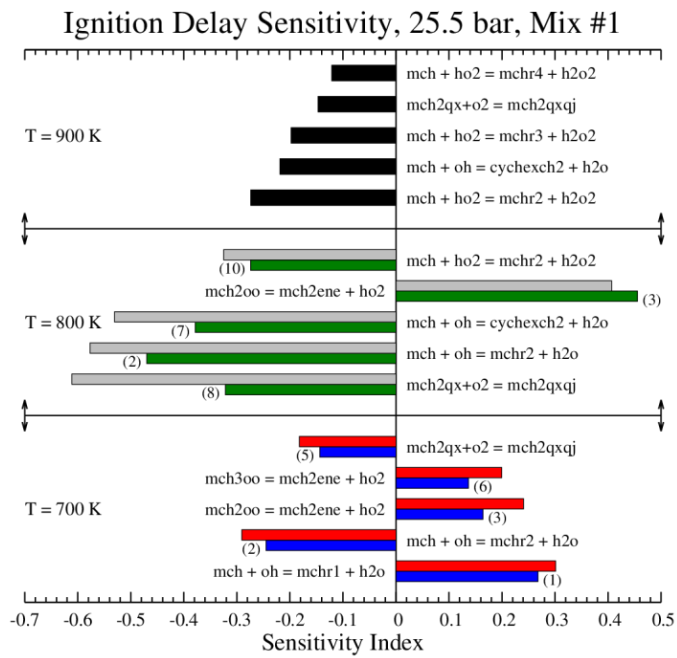

Figure 11: Sensitivity of the ignition delay to various reaction rates for Mix #1 ($\phi$=1.0), 25.5 bar and three temperatures (700 K, 800 K, and 900 K). At 700 K, the sensitivity of the overall ignition delay is in red and the sensitivity of the first stage ignition delay is in blue. At 800 K, the sensitivity of the overall ignition delay is in grey and the sensitivity of the first stage ignition delay is in green. At 900 K, the sensitivity of the overall ignition delay is in black. Numbers in parentheses represent the ranking of the first stage sensitivity indices.

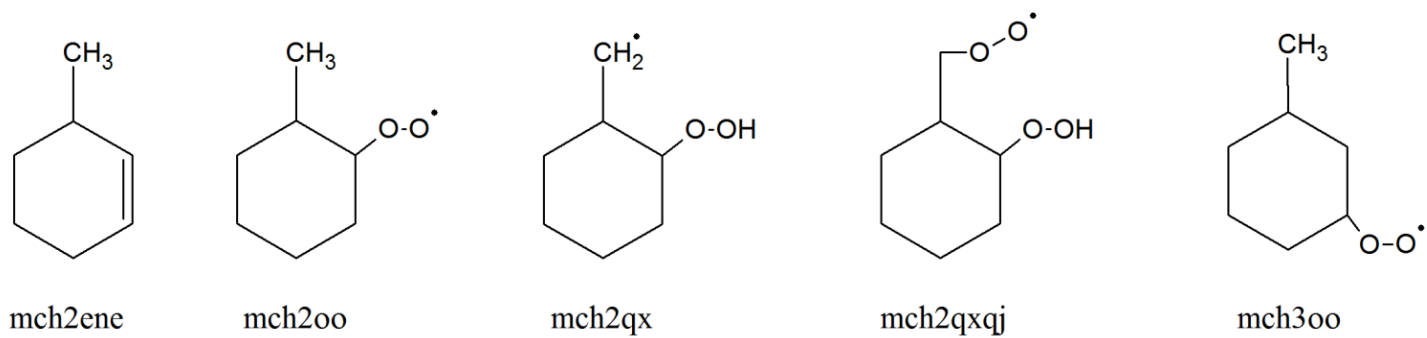

Figure 12: Species mentioned in Figure 11 or Table 2 and not included in Figure 10.

**Table 2: Reactions that eliminate the first inflection point for a nominal case with two-stage ignition.**

| Multiplication | Division |
|---|---|
| mch2oo = mch2ene + ho2 | mch2qx + o2 = mch2qxqj |
| mch3oo = mch2ene + ho2 | |
| mch3oo = mch3ene + ho2 | |
| mch + oh = mchr1 + h2o | |
| mch + oh = mchr4 + h2o | |
| mch + oh = mchr3 + h2o | |

In general, Fig. 11 shows that the ignition delay is sensitive to different sets of reactions at the three temperatures, although there is some overlap. The overlapping reactions confound simple recommendations for rate improvements. For instance, at 700 K, increasing the overall ignition delay will improve agreement with the experimental data, but at 800 K, the agreement is already quite good. Therefore, adjusting any of the rates to improve the agreement with the overall ignition delay at 700 K will probably make the agreement worse at 800 K. However, the first stage ignition delays at 700 K and 800 K are both under-predicted; furthermore, two-stage ignition is predicted at temperatures for which the experimental ignition is single stage. It should therefore be possible to adjust several rate constants simultaneously to improve agreement with the first stage ignition delay and not deteriorate agreement with the overall ignition delay. To accomplish the simultaneous improvement of agreement of first stage and overall ignition delays, the rate constants of reactions that control the second stage ignition delay may also need to be adjusted (where second stage ignition delay is the difference between the overall ignition delay and the first stage ignition delay).

Interestingly, the formation and destruction reactions of ROOH species do not appear in Fig. 11, despite their importance in the destruction of ROO radicals, particularly at 700 K (see Fig. 10). This may be due to the fact that formation of ROOH by reaction with $HO_2$ followed by consumption of ROOH is a chain propagation path through the reactions $ROO + HO_2 = ROOH + O_2$; $ROOH = RO + OH$. In this sequence two radicals are formed (RO, OH) and two radicals are consumed (ROO, $HO_2$). Thus, the formation of ROOH by reaction with $HO_2$ and its subsequent destruction has a somewhat neutral effect on the radical pool.

At 900 K, the overall ignition delay is particularly sensitive to reactions that form hydrogen peroxide, which decomposes to two hydroxyl radicals as the temperature increases during the induction period. Therefore, increasing the rate of formation of hydrogen peroxide will increase the formation of hydroxyl radical and decrease the overall ignition delay. At 900 K, the overall ignition delay is over-predicted, so to improve the results the overall ignition delay should be reduced (i.e. increasing the rates of reactions with negative sensitivity will improve the comparison). In addition, many of the reactions that are important at 900 K are not important at 700 and 800 K, implying that changes made to the rates to improve the high-temperature agreement will not significantly change the agreement at lower temperature. In particular, the MCH+HO$_2$ rate constants have not been measured or calculated to our knowledge and are based on acyclic alkane rate constants [42]. They have uncertainties of at least a factor of 2 and as much as a factor of 10 based on the work of Aguilera-Iparraguirre et al. [42]. Increasing these rate constants would improve the agreement with the experimental ignition data at 900 K in the RCM and shock tube. Experimental measurements and theoretical calculations are needed for the fuel+HO$_2$ reaction class to reduce this uncertainty in the rate constants.

## 6. Conclusion

In this study, new experimental data are collected for methylcyclohexane autoignition in a heated rapid compression machine. Following the work of Mittal and Sung [16], three mixtures of MCH/O$_2$/N$_2$/Ar at equivalence ratios of $\phi$=0.5, 1.0, and 1.5 are used and the ignition delays are measured at compressed pressure of 50 bar, for compressed temperatures in the range of 690–900 K. Two-stage ignition phenomena are reported for the stoichiometric and rich mixtures. However, substantial reactivity during the compression stroke limited the temperature range over which ignition delays could be reported, especially for the lean case. For these mixtures where the fuel concentration was kept constant, the order of reactivity, in terms of inverse overall ignition delay, is $\phi$=0.5> $\phi$=1.0> $\phi$=1.5.

In addition, an existing model for the combustion of MCH developed by Pitz et al. [15] is updated with new reaction rates and pathways. The new model shows good agreement with the overall ignition delays measured in this study, as well as the overall ignition delays measured in the studies of Mittal and Sung [16], Vasu et al. [9], and Vanderover and Oehlschlaeger [10]. However, the first stage ignition delays are uniformly under-predicted and in several cases, first stage ignition is predicted by the model where experimental ignition response shows no two-stage character. To help understand the fuel decomposition pathways and the reactions controlling the ignition, further analysis of the present mechanism is conducted.

First, reaction path analysis is conducted for low-, intermediate-, and high-temperature ignition considered in this study. The results show that MCH primarily decomposes by H-abstraction reactions involving OH and $HO_2$ radicals, followed by oxygen addition reactions. At low temperatures, the oxygen addition is followed by isomerization to QOOH species and second oxygen addition, leading to the low-temperature chain branching characteristic of two-stage ignition. At intermediate temperatures, the elimination of methylcyclohexene and $HO_2$ becomes competitive with the isomerization reaction, leading to the NTC region of the overall ignition delay. Finally, at high temperatures, $HO_2$+MCH reactions form $H_2O_2$ and end the NTC region.

Second, a brute force sensitivity analysis is conducted to identify the reactions of the fuel and primary fuel radicals that control the ignition process. The overall and first stage ignition events at low and intermediate temperatures are primarily controlled by the initial reactions to form fuel radicals, especially H-abstraction by OH. At high temperatures, the controlling reactions are still the fuel radical formation reactions, but now the ignition process is controlled by H-abstraction by hydroperoxyl instead of hydroxyl. Combined, these analyses suggest that further investigation of several of the low-temperature fuel decomposition pathways is required and more accurate rate constants for fuel+$HO_2$ reactions are needed.

**Acknowledgements**

The work at the University of Connecticut was supported as part of the Combustion Energy Frontier Research Center, an Energy Frontier Research Center funded by the US Department of Energy, Office of Science, Office of Basic Energy Sciences, under Award Number DE-SC0001198. The work at LLNL was supported by U.S. Department of Energy, Office of Vehicle Technologies, program manager Gurpreet Singh and performed under the auspices of the U.S. Department of Energy by Lawrence Livermore National Laboratory under Contract No. DE-AC52-07NA27344. Alexander Davis acknowledges funding from KAUST CCRC with technical monitoring of Dr. Mani Sarathy.

**References**

[1]     Y. Briker, Z. Ring, A. Iacchelli, N. McLean, P.M. Rahimi, C. Fairbridge, R. Malhotra, M.A. Coggiola, S.E. Young, Energ. Fuel 15 (2001) 23–37.

[2]     W.J. Pitz, N.P. Cernansky, F.L. Dryer, F.N. Egolfopoulos, J.T. Farrell, D.G. Friend, H. Pitsch, Development of an experimental database and chemical kinetic models for surrogate gasoline fuels, Report No. 2007-01-0175, SAE International, 2007.

[3]     J.T. Farrell, N.P. Cernansky, F.L. Dryer, D.G. Friend, C.A. Hergart, C.K. Law, R.M. McDavid, C.J. Mueller, A.K. Patel, H. Pitsch, Development of an Experimental Database and Kinetic Models for Surrogate Diesel Fuels, Report No. 2007-01-0201, SAE International, 2007.

[4]     M.B. Colket, J.T. Edwards, S. Williams, N.P. Cernansky, D.L. Miller, F.N. Egolfopoulos, P. Lindstedt, K. Seshadri, F.L. Dryer, C.K. Law, D.G. Friend, D.B. Lenhert, H. Pitsch, A. Sarofim, M.D. Smooke, W. Tsang, Development of an experimental database and kinetic models for surrogate jet fuels, 45th AIAA Aerosp. Sci. Meet., American Institute of Aeronautics and Astronautics, 2007.

[5]     T. Bieleveld, A. Frassoldati, A. Cuoci, T. Faravelli, E. Ranzi, U. Niemann, K. Seshadri, Proc. Combust. Inst. 32 (2009) 493–500.

[6]     C. V. Naik, W.J. Pitz, C.K. Westbrook, M. Sjöberg, J.E. Dec, J.P. Orme, H.J. Curran, J.M. Simmie, Detailed Chemical Kinetic Modeling of Surrogate Fuels for Gasoline and Application to an HCCI Engine, Report No. 2005-01-3741, SAE International, 2005.

[7]     W.J. Pitz, C.J. Mueller, Prog. Energy Combust. Sci. 37 (2011) 330–350.

[8]     B. Rotavera, E.L. Petersen, Proc. Combust. Inst. 34 (2013) 435–442.

[9]     S.S. Vasu, D.F. Davidson, Z. Hong, R.K. Hanson, Energ. Fuel 23 (2009) 175–185.

[10]    J. Vanderover, M.A. Oehlschlaeger, Int. J. Chem. Kinet. 41 (2009) 82–91.

[11]    R.D. Hawthorn, A.C. Nixon, AIAA J. 4 (1966) 513–520.

[12]    J.P. Orme, H.J. Curran, J.M. Simmie, J. Phys. Chem. A 110 (2006) 114–131.

[13]    Z. Hong, K.-Y. Lam, D.F. Davidson, R.K. Hanson, Combust. Flame 158 (2011) 1456–1468.

[14]    S. Tanaka, F. Ayala, J.C. Keck, J.B. Heywood, Combust. Flame 132 (2003) 219–239.

[15]    W.J. Pitz, C. V. Naik, T.N. Mhaoldúin, C.K. Westbrook, H.J. Curran, J.P. Orme, J.M. Simmie, Proc. Combust. Inst. 31 (2007) 267–275.

[16] G. Mittal, C.-J. Sung, Combust. Flame 156 (2009) 1852–1855.

[17] Y. Yang, A.L. Boehman, J.M. Simmie, Combust. Flame 157 (2010) 2357–2368.

[18] J.A. Montgomery, M.J. Frisch, J.W. Ochterski, G.A. Petersson, J. Chem. Phys. 112 (2000) 6532.

[19] M.J. Frisch, G.W. Trucks, H.B. Schlegel, G.E. Scuseria, M.A. Robb, J.R. Cheeseman, G. Scalmani, V. Barone, B. Mennucci, G.A. Petersson, H. Nakatsuji, M. Caricato, X. Li, H.P. Hratchian, A.F. Izmaylov, J. Bloino, G. Zheng, J.L. Sonnenberg, M. Hada, M. Ehara, K. Toyota, R. Fukuda, J. Hasegawa, M. Ishida, T. Nakajima, Y. Honda, O. Kitao, H. Nakai, T. Vreven, J.A. Montgomery Jr., J.E. Peralta, F. Ogliaro, M. Bearpark, J.J. Heyd, E. Brothers, K.N. Kudin, V.N. Staroverov, R. Kobayashi, J. Normand, K. Raghavachari, A. Rendell, J.C. Burant, S.S. Iyengar, J. Tomasi, M. Cossi, N. Rega, J.M. Millam, M. Klene, J.E. Knox, J.B. Cross, V. Bakken, C. Adamo, J. Jaramillo, R. Gomperts, R.E. Stratmann, O. Yazyev, A.J. Austin, R. Cammi, C. Pomelli, J.W. Ochterski, R.L. Martin, K. Morokuma, V.G. Zakrzewski, G.A. Voth, P. Salvador, J.J. Dannenberg, S. Dapprich, A.D. Daniels, Ö. Farkas, J.B. Foresman, J. V Ortiz, J. Cioslowski, D.J. Fox, Gaussian 09 Revision A.1, Gaussian Inc., Wallingford, CT, 2009.

[20] V. Mokrushin, V. Bedanov, W. Tsang, M. Zachariah, V. Knyazev, ChemRate 1.5.8, NIST, Gaithersburg, MD, 2009.

[21] K.S. Pitzer, W.D. Gwinn, J. Chem. Phys. 10 (1942) 428–440.

[22] B.W. Weber, K. Kumar, Y. Zhang, C.-J. Sung, Combust. Flame 158 (2011) 809–819.

[23] K. Kumar, G. Mittal, C.-J. Sung, Combust. Flame 156 (2009) 1278–1288.

[24] A.K. Das, C.-J. Sung, Y. Zhang, G. Mittal, Int. J. Hydrogen Energy 37 (2012) 6901–6911.

[25] G. Mittal, C.-J. Sung, Combust. Sci. Technol. 179 (2007) 497–530.

[26] CHEMKIN-PRO 15131, Reaction Design: San Diego, 2013.

[27] W.K. Metcalfe, S.M. Burke, S.S. Ahmed, H.J. Curran, Int. J. Chem. Kinet. (2013) DOI: 10.1002/kin.20802.

[28] H. Nakamura, D. Darcy, M. Mehl, C.J. Tobin, W.K. Metcalfe, W.J. Pitz, C.K. Westbrook, H.J. Curran, Combust. Flame (2013) DOI: 10.1016/j.combustflame.2013.08.002.

[29] E.J. Silke, W.J. Pitz, C.K. Westbrook, M. Ribaucour, J. Phys. Chem. A 111 (2007) 3761–3775.

[30] R. Sivaramakrishnan, J.V. Michael, Combust. Flame 156 (2009) 1126–1134.

[31] S.M. Sarathy, C.K. Westbrook, M. Mehl, W.J. Pitz, C. Togbé, P. Dagaut, H. Wang, M.A. Oehlschlaeger, U. Niemann, K. Seshadri, P.S. Veloo, C. Ji, F.N. Egolfopoulos, T. Lu, Combust. Flame 158 (2011) 2338–2357.

[32] W.J. Pitz, S.A. Skeen, M. Mehl, N. Hansen, E.J. Silke, Chemical Kinetic Modeling of Low Pressure Methycyclohexane Flames, 8th US Natl. Combust. Meet., The Combustion Institute, 2013.

[33] R.X. Fernandes, J. Zádor, L.E. Jusinski, J.A. Miller, C.A. Taatjes, Phys. Chem. Chem. Phys. 11 (2009) 1320–1327.

[34] S.M. Villano, L.K. Huynh, H.-H. Carstensen, A.M. Dean, J. Phys. Chem. A 116 (2012) 5068–5089.

[35] S.K. Gulati, R.W. Walker, J. Chem. Soc. Faraday Trans. 2 85 (1989) 1799–1812.

[36] B. Husson, O. Herbinet, P.A. Glaude, S.S. Ahmed, F. Battin-Leclerc, J. Phys. Chem. A 116 (2012) 5100–5111.

[37] E.R. Ritter, J.W. Bozzelli, Int. J. Chem. Kinet. 23 (1991) 767–778.

[38] S. Sharma, S. Raman, W.H. Green, J. Phys. Chem. A 114 (2010) 5689–5701.

[39] K. Kumar, C.-J. Sung, Combust. Flame 157 (2010) 676–685.

[40] H.J. Curran, P. Gaffuri, W.J. Pitz, C.K. Westbrook, Combust. Flame 129 (2002) 253–280.

[41] J. Zádor, C.A. Taatjes, R.X. Fernandes, Prog. Energy Combust. Sci. 37 (2011) 371–421.

[42] J. Aguilera-Iparraguirre, H.J. Curran, W. Klopper, J.M. Simmie, J. Phys. Chem. A 112 (2008) 7047–7054.

**Figure Captions**

Figure 1: Representative pressure trace indicating the definition of the first stage ($\tau_1$) and overall ($\tau$) ignition delays and the corresponding non-reactive pressure trace. EOC stands for End of Compression.

Figure 2: Reaction path diagram for the 1,5 H-migration reaction in 2-methylcyclohexyl-1-peroxy (c2McHp) radical at 298 K. Energies are in kJ/mol. Blue dashed line represents pathway described in Yang et al. [17].

Figure 3: Experimentally measured ignition delays at $P_C$=50 bar for the mixture conditions in Table 1.

Figure 4: Comparison of experimental and simulated ignition delays for three pressures for Mix #1. The data at 15.1 and 25.5 bar are from the study of Mittal and Sung [16]. (a) First stage ignition delays (b) Overall ignition delays.

Figure 5: Comparison of experimental and simulated ignition delays for three pressures for Mix #2. The data at 15.1 and 25.5 bar are from the study of Mittal and Sung [16]. (a) First stage ignition delays (b) Overall ignition delays.

Figure 6: Comparison of experimental and simulated ignition delays for three pressures for Mix #3. The data at 15.1 and 25.5 bar are from the study of Mittal and Sung [16]. (a) First stage ignition delays (b) Overall ignition delays.

Figure 7: Comparison of selected simulated and experimental pressure traces at $P_C$=50 bar for (a) Mix #1 (b) Mix #2 (c) Mix #3. Red lines indicate that the pressure profile of the reactive simulation deviates from the non-reactive case prior to EOC. Solid lines: experiment; dashed lines: reactive simulation; dot-dot-dashed lines: non-reactive simulation.

Figure 8: Comparison of the present model with the experiments from Vasu et al. [9] and Vanderover and Oehlschlaeger [10] near 50 atm and for stoichiometric mixtures in $O_2/N_2$ air.

Figure 9: Comparison of mechanism performance with the activation energy of Reaction Class 28, ketohydroperoxide decomposition, set at 41.6 kcal/mol (blue) and 39.0 kcal/mol (red). Experimental ignition delays are shown in green symbols.

Figure 10: Path analysis of MCH combustion. Initial conditions are 25.5 bar and Mix #1 ($\phi$=1.0) and 700 K (plain text), 800 K (bold text), 900 K (italic text). Note that not all possible reaction pathways are shown for each species.

Figure 11: Sensitivity of the ignition delay to various reaction rates for Mix #1 ($\phi$=1.0), 25.5 bar and three temperatures (700 K, 800 K, and 900 K). At 700 K, the sensitivity of the overall ignition delay is in red and the sensitivity of the first stage ignition delay is in blue. At 800 K, the sensitivity of the overall ignition delay is in grey and the sensitivity of the first stage ignition delay is in green. At 900 K, the sensitivity of the overall ignition delay is in black. Numbers in parentheses represent the ranking of the first stage sensitivity indices.

Figure 12: Species mentioned in Figure 11 or Table 2 and not included in Figure 10.

**Table Captions**

**Table 1: Molar Proportions of Reactants**

**Table 2: Reactions that eliminate the first inflection point for a nominal case with two-stage ignition.**